\documentclass[%
 reprint,
 amsmath,amssymb,
 aps,
]{revtex4-2}

\usepackage{graphicx}
\usepackage{dcolumn}
\usepackage{bm}

\begin{document}

\preprint{APS/123-QED}

\title{TemplateGeNN: Neural Networks used to accelerate Gravitational Wave Template Bank Generation}

\author{Susanna Green}
\affiliation{Institute of Cosmology and Gravitation, University of Portsmouth, Portsmouth PO1 3FX, United Kingdom}

\author{Andrew Lundgren}
\affiliation{Catalan Institution for Research and Advanced Studies (ICREA), E-08010 Barcelona, Spain}
\affiliation{Institut de F\'isica d’Altes Energies (IFAE), The Barcelona Institute
of Science and Technology, UAB Campus, E-08193 Barcelona, Spain}

\date{\today}

\begin{abstract}
We introduce TemplateGeNN, a fast stochastic template bank generation algorithm which uses Graphical Processing Units (GPUs) and a LearningMatch model (Siamese neural network). TemplateGeNN generated a binary black hole template bank (chirp mass varied from $5 M_{\odot} \leq \mathcal{M}_{c} \leq 20M_{\odot}$, symmetric mass ratio varied from $0.1 \leq \eta \leq 0.24999$, and equal aligned spin varied from $-0.99 \leq \chi_{1,2}\leq 0.99$) of 31,640 templates in $\sim 1$ day on a single A100 GPU. To test the sensitivity of this template bank we injected 7746 binary black hole templates into LIGO Gaussian noise. This template bank recovered 98$\%$ of the injections with a fitting factor greater than 0.97. For lower mass regions (black hole mass region between $5 M_{\odot} \leq m_{1, 2} \leq 25 M_{\odot}$), 99$\%$ of 9469 injections were recovered with a fitting factor greater than 0.97. LearningMatch and TemplateGeNN are a machine-learning pipeline that can be used to accelerate template bank generation for future gravitational-wave data analysis. 
\end{abstract}

\maketitle

\section{Introduction}
In 2015, the first gravitational wave from a binary black hole merger was detected by the Laser Interferometer Gravitational-wave Observatory, otherwise known as LIGO \citep{GW150914}. Since then, Advanced LIGO \citep{AdvancedLIGO2015} and Advanced Virgo \citep{AdvancedVirgo2015} (an interferometer hosted by the European Gravitational Observatory) have observed many binary black hole mergers and even a binary neutron star merger during the first two observing runs \citep{GWTC1, GW170817}. The third observing run, which KAGRA \citep{Akutsu2019} also joined, led to the detection of 74 gravitational-wave events including the observation of the first asymmetric binary black hole merger, the first inspiral of intermediate-mass black holes, and proposed neutron star-black hole candidates \citep{Abbott2020ProspectsKAGRA, GWTC2, GWTC2.1, GWTC3, GW190412, GW190521, GW190814}. More gravitational-wave events have been observed during the fourth observing run which started in May 2023 \citep{Abbott2020ProspectsKAGRA, 2024arXiv241114607C, 2024arXiv240902831S}. However, identifying these cosmic events is computationally expensive. As the sensitivity of gravitational-wave detectors is increased and more detectors join the network, fast data analysis will be essential to make future gravitational-wave observations a success.

The gravitational waves observed in the past observing runs were identified by various gravitational-wave search pipelines. Modelled searches, specifically PyCBC \citep{PyCBCSearch2016, Nitz2018RapidLive}, Multi-band template analysis (MBTA) \citep{Adams2016, Adams2015LowMBTA, Aubin2021}, GstLAL \citep{Cannon2021GstLAL:Discovery, Ewing2024PerformanceRun, Mukherjee2021TemplateVirgo, Sakon2024TemplateKAGRA}, and Summed Parallel Infinite Impulse Response (SPIIR) \citep{Kovalam2022EarlySearch, Chu2022}, use matched-filtering to identify these gravitational waves. Matched-filtering is a signal processing technique that is used to extract a known signal out of noisy data. These search pipelines identify a gravitational wave by matching a large number of templates, waveform models that approximate the actual gravitational wave signal, with the data. The template which produces the largest signal-to-noise ratio is the preferred template. Template banks are a catalogue of templates required to cover the parameter space sufficiently so that a gravitational-wave signal will not be missed. Template banks are also required to not over-cover the parameter space so that unnecessary calculations are not performed \citep{Allen2021OptimalBanks}. As a consequence, various template bank generation techniques have been proposed in the literature to meet these requirements but many of them are computationally expensive. Geometric (or lattice) algorithms \citep{Roulet2019TemplateAlgorithm, Cokelaer2007GravitationalSignals, Prix2007Template-basedSpaces, Hanna2023BinaryMergers}, stochastic (or random) algorithms \citep{Harry2009, Allen2022PerformanceBanks, Fehrmann2014EfficientAlgorithm}, and a combination between the two \citep{Roy2017, Roy2019EffectualData} have been all been used to generate the template banks. In recent years there have been attempts to introduce novel methods to speed up these algorithms such as normalizing flows \citep{Schmidt2024Gravitational-waveBinaries}. 

In this paper, a Siamese neural network, called LearningMatch, is integrated into a stochastic template bank algorithm and Graphical Processing Units (GPUs) are used to generate a template bank quickly. For this paper, it is assumed that the reader is familiar with LearningMatch, see reference for more information \citep{Green2025LearningMatch}. In section \ref{Theory}, template bank generation is discussed specifically focusing on the stochastic template bank generation algorithm. The results are presented in section \ref{Results} and then discussed in section \ref{Discussion} showcasing that TemplateGeNN can produce a template bank that could be used by the LIGO-Virgo-KAGRA (LVK) Collaboration. Finally, in section \ref{Conclusions} the final comments are made. All the code is open source and can be accessed at \url{https://github.com/SusannaGreen/TemplateGeNN}.

\section{Template Bank Generation}
\label{Theory}

The aim of a template bank is to sufficiently cover the search parameter space; be sparse enough that unnecessary calculations are not performed, but populated enough that a gravitational-wave signal will not be missed. There are two main types of template placement algorithms that are used to search for compact binary coalescence: geometric and stochastic methods. In the geometric method the templates are placed in an hexagonal lattice using the \textit{chirp times}, with the first order contribution defined as
\begin{equation}
    \tau_{0} = \frac{5}{256\pi \eta f_{L}}(\pi M f_{L})^{-5/3},
\end{equation}
where $f_{L}$ is the lower-cut-off frequency, $\eta$ is the symmetric mass ratio ($(m_{1}m_{2})(m_{1}+m_{2})^{-2}$), and \textit{M} denotes the total mass
of the binary \citep{Owen1999, Harry2014}. \textit{Chirp times} are used rather than the standard component masses because the signal manifold at 1PN order is flat, and therefore the \textit{chirp times} are Cartesian-like coordinates \citep{Sathyaprakash1994, Babak2006}. In higher dimensions, determining the geometric placement is challenging due to the potential absence of a known analytical metric and the uncertainty surrounding the optimal placement in curved spaces \citep{Roy2017, Brown2012}. An alternative algorithm is used called the stochastic method, which builds the template bank iteratively \citep{Harry2009}. Firstly, an arbitrary template in the parameter space is chosen and a new template is proposed, the match between these two templates is calculated and is defined as the inner product maximised over the time of coalescence and phase of the waveform and can be expressed as
\begin{equation}\label{match}
\begin{split}
    \mathcal{M}(h_{1}, h_{2}) &= \max_{\phi_{c}, t_{c}} (h_{1}|h_{2}) \\
    &= \max_{\phi_{c}, t_{c}} \left[ 4 \Re \int_{0}^{\infty} \frac{\tilde{h}_{1}^{*}(f)\tilde{h}_{2}(f)}{S_{h}(\textit{f})} df
    \right]
\end{split}
\end{equation}
were $h_{1}$ and $h_{2}$ are two templates with unit norm and $S_{n}(f)$ is the one-sided noise power spectral density (PSD) \citep{Brown2012}. $S_{n}(f)$ is defined as, 

\begin{equation}\label{average noise}
\langle \tilde{n}^{*}(f) \tilde{n}(f') \rangle = \delta(f-f')\frac{1}{2}{S_{n}(\textit{f})},
\end{equation}

where \textit{n(f)} is the noise spectrum in the frequency domain and mathematical symbols have their standard meaning. If the match is less than the desired minimal match, which for compact binary coalescence template banks is normally chosen to be 0.97, then the proposed template is added to the template bank otherwise it is rejected. The value of the minimal match is chosen so that a maximum $3\%$ of the signal-to-noise ratio is lost when the template bank recovers an actual gravitational-wave signal and this value is known as the fitting factor \citep{Owen1995, Apostolatos1995}. Once the template bank is generated, the template bank needs to be verified (i.e. confirm there is no gaps in the coverage) so that a gravitational wave signal will not be missed by the LVK Collaboration.

\section{Results}
\label{Results}

\begin{figure*}
    \centering
    \includegraphics[scale=0.4]{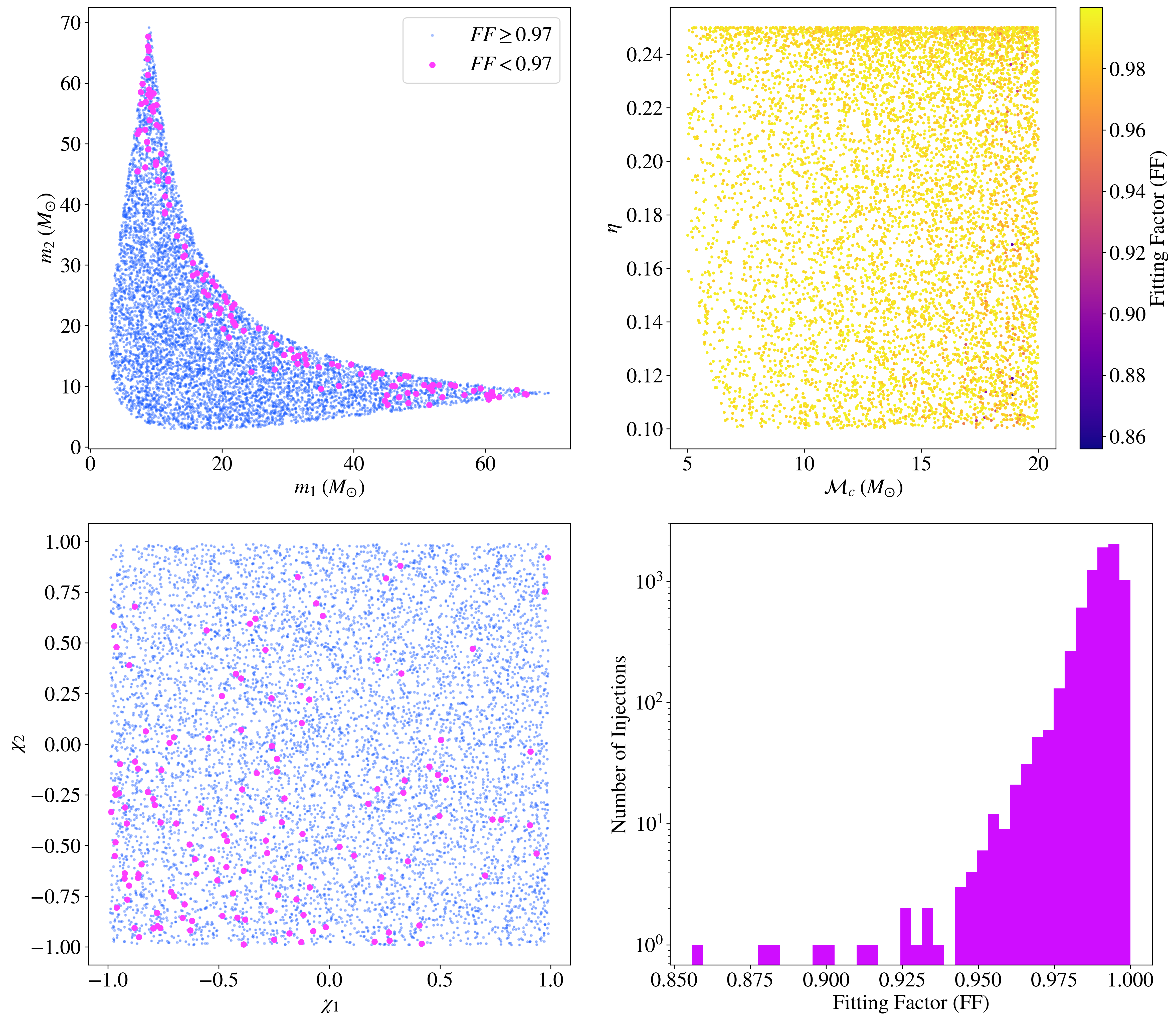}
    \caption{7746 binary black hole mergers were injected into Gaussian noise and the template bank generated by TemplateGeNN was tested. The top left figure shows the mass of the injected signals and the bottom left shows the spin of the injected signals. Both figures showcase 130 binary black hole injections that had fitting factors less than 0.97 which means that these injections were not recovered by the template bank. The top right figure shows the fitting factors calculated for all 7746 signals in the chirp mass and symmetric mass ratio parameter space. The bottom right figure is a histogram of fitting factors produced by the \textit{PyCBC template bank verificator}.}
    \label{fig: Injections Big}
\end{figure*}

\begin{figure*}
    \centering
    \includegraphics[scale=0.4]{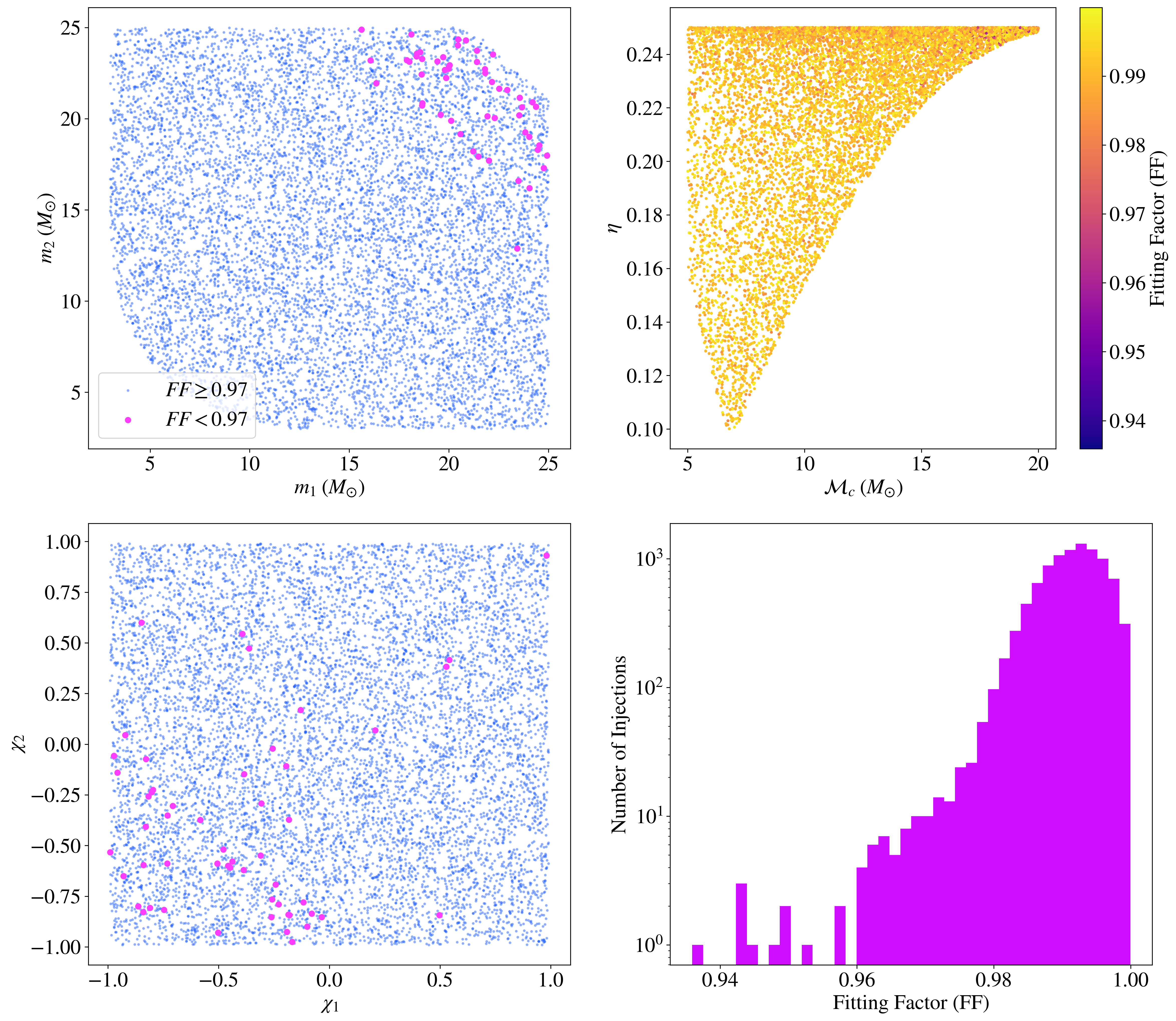}
    \caption{9469 binary black hole mergers were injected into Gaussian noise and the lower mass regions (black hole mass region between $5 M_{\odot} \leq m_{1, 2} \leq 25 M_{\odot}$) of the template bank were tested. The top left figure shows the mass of the injected signals and the bottom left shows the spin of the injected signals. Both figures showcase 53 binary black hole injections that had fitting factors less than 0.97 which means that these injections were not recovered by the template bank. The top right figure shows the fitting factors calculated for all 9469 signals in the chirp mass and symmetric mass ratio parameter space. The bottom right figure is a histogram of fitting factors produced by the \textit{PyCBC template bank verificator}.}
    \label{fig: Injections Small}
\end{figure*}

\begin{figure}
    \centering
    \includegraphics[scale=0.53]{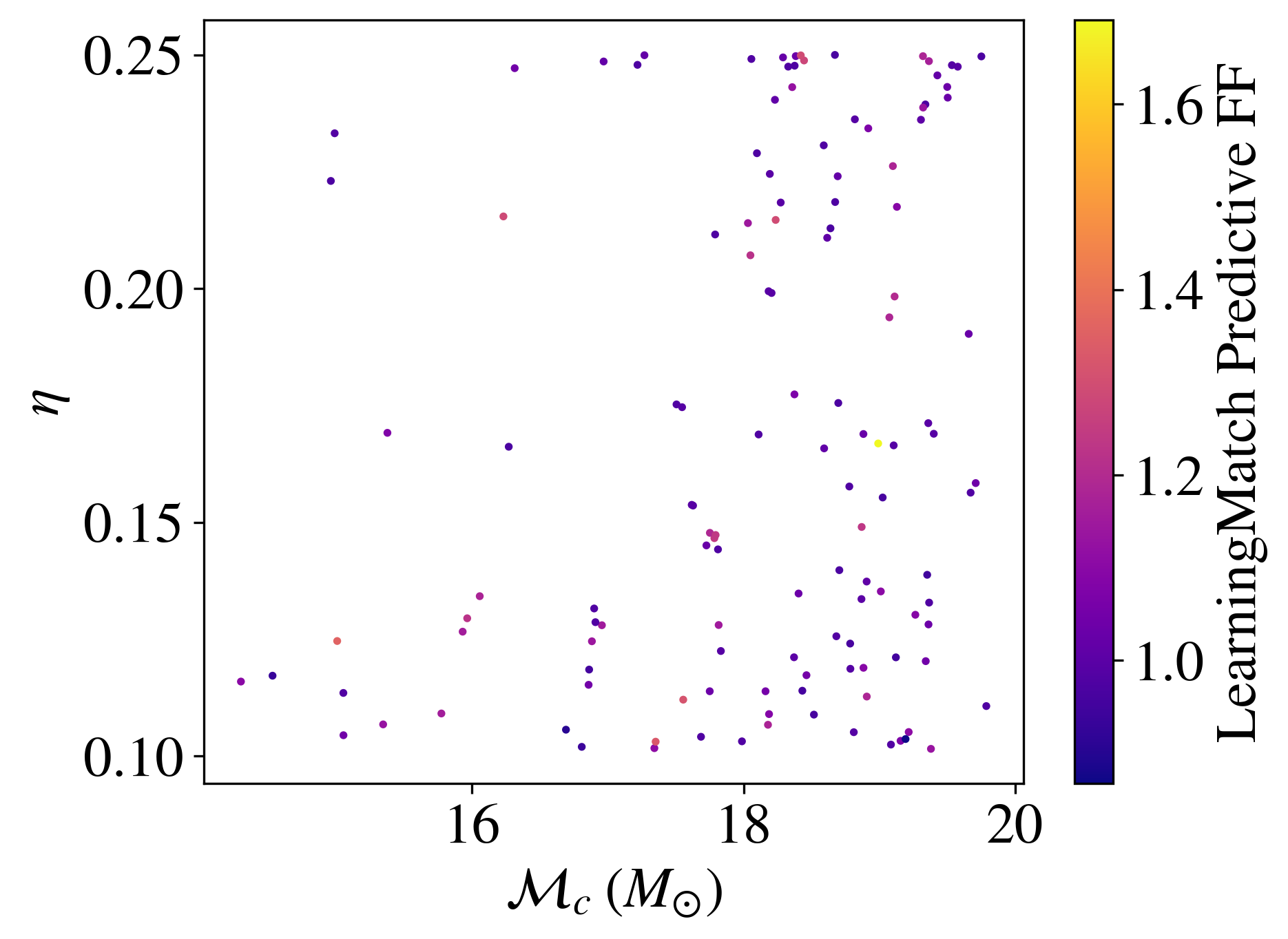}
    \caption{The \textit{LearningMatch Predictive Fitting Factor (FF)} was computed for the 130 binary black hole templates that were injected into Gaussian noise and were not recovered by the template bank generated by TemplateGeNN. \textit{LearningMatch Predictive FF} is defined as the maximum match (i.e. fitting factor) predicted by the LearningMatch between the template bank and the missed binary black hole merger injections.}
    \label{fig: LearningMatch Fitting Factor}
\end{figure}

A trained LearningMatch model is integrated into a stochastic template bank generation algorithm and then run on GPUs. LearningMatch is a Siamese neural network that predicts the match between two gravitational-wave templates given their mass and spin parameters \cite{Green2025LearningMatch}. A binary black hole template bank of 31,640 templates can be generated in approximately 1 day on a single A100 GPU. TemplateGeNN was considered to have converged when only 1 out of 5000 suggested templates was accepted into the bank. The template bank was generated in the $\lambda_{0}$ ($\lambda_{0}\propto \mathcal{M}_{c}^{-5/3}$), symmetric mass ratio ($\eta$) and equal aligned spin ($\chi_{1}$ = $\chi_{2}$) parameter space. This parametrisation is due to the LearningMatch model and was originally chosen because the template bank is expected to have a roughly constant density in $\lambda_{0}$. A template bank was generated using $\lambda_{0}$ that varied between 0.006 and 1.4. This corresponds to a $\mathcal{M}_{c}$ (chirp mass) range that varies between $5 M_{\odot}$ and $20 M_{\odot}$. $\mathcal{M}_{c}$ is expressed as $\mathcal{M}_{c} = (m_{1}m_{2})^{3/5}(m_{1}+m_{2})^{-1/5}$, where $m_{1}$ and $m_{2}$ are the mass of the two black holes. $\eta$ is expressed as $\eta = (m_{1}m_{2})(m_{1}+m_{2})^{-2}$ and varied between 0.1 and 0.24999 ($\eta=0.25$ was not used to avoid numerical failure in the waveform). The spin of both black holes varied between -0.99 and 0.99 \citep{Roy2019EffectualData}. Using \textit{PyCBC template bank verificator} \citep{PyCBCSearch2016}, binary black hole mergers were injected into Gaussian noise to determine how well this template bank covered this parameter space. The binary black hole mergers were modelled using the `IMRPhenomXAS' waveform \citep{IMRPhenomXAS}. A simulated LIGO PSD representative of the third observing run was used with a sampling frequency of 2048 Hz and a low-frequency cut-off of 15 Hz (called \textit{aLIGO140MpcT1800545} in PyCBC) \citep{PyCBCPSD}. 7746 binary black hole mergers were injected into the entire parameter space while 9469 binary black hole mergers were injected in the lower mass range (specifically focusing on the black hole mass region between $m_{1, 2}= 5 M_{\odot} - 25 M_{\odot}$). Figure \ref{fig: Injections Big} shows that $98\%$ (7316 out of 7746) of binary black hole injections have a fitting factor greater than 0.97 and therefore would be recovered by this template bank. Figure \ref{fig: Injections Small} focuses on the low mass regions of the template bank and shows that $99\%$ (9416 out of 9469) of binary black hole injections have a fitting factor greater than 0.97 and therefore would be recovered by the template bank. From both Figure \ref{fig: Injections Big} and Figure \ref{fig: Injections Small}, we noticed that the binary black hole merger injections that are not recovered by the template bank were in the high mass regions. LearningMatch was then used to determine the maximum match (i.e. fitting factor) between the template bank and the missed binary black hole merger injections. In Figure \ref{fig: LearningMatch Fitting Factor}, we noticed that LearningMatch had predicted match values greater than 1 which is not physically possible. Therefore, we conclude that LearningMatch was responsible for the template bank missing these binary black hole injections.

\section{Discussion}
\label{Discussion}

TemplateGeNN was able to generate a template bank in $\sim 1$ day using a single A100 GPU. The speed-up observed in TemplateGeNN is because the template bank is generated on GPUs rather than CPUs. Current template bank generation pipelines use CPU hardware to compute the match and generate the template. It is worth mentioning that this may not always be the case because there is research making template generation GPU compatible using JAX, see reference for more information \citep{Edwards2023Ripple:Analysis}. 

\textit{PyCBC template bank verificator} was used to determine how well the template bank generated by TemplateGeNN would recover binary black hole mergers injected into Gaussian noise. Out of 7746 injections, 7316 of injections were recovered by the template bank with a fitting factor greater than 0.97, as shown in Figure \ref{fig: Injections Big}. We then focused on the lower mass regions of the template bank and out of 9469 injections, 9416 were recovered with a fitting factor greater than 0.97, as seen in Figure \ref{fig: Injections Small}. In both Figures \ref{fig: Injections Big} and \ref{fig: Injections Small}, the injections that were not recovered by the template bank (i.e. fitting factors less than 0.97) were all in the high mass regions. 

For all the black hole merger injections that were not recovered, LearningMatch was used to determine the maximum match (i.e. fitting factor) between the template bank and the injection, we called this parameter the \textit{LearningMatch Predictive Fitting Factor}. In Figure \ref{fig: LearningMatch Fitting Factor}, \textit{LearningMatch Predictive Fitting Factor} had values greater than 1, which means LearningMatch had predicted match values greater than 1 which is not physically possible. Therefore, we concluded that LearningMatch was responsible for the template bank missing these binary black hole injections. The reason that LearningMatch failed to learn the high mass regions is due to the parametrisation. LearningMatch was trained using the $\lambda_{0}$ parametrisation, compared to the $m_{1}$ and $m_{2}$ parametrisation, because this results in the templates being uniformly distributed in this parameter space. Consequently, in the lower mass regions, the templates get stretched out in $\lambda_{0}$ while in the higher mass regions, the templates get squashed. Therefore, when $\lambda_{0}$ values are taken from a uniform distribution (which is how the training datasets for LearningMatch were created), the higher mass regions are under-covered with data points. This behaviour was also discussed in the original paper (see reference \citep{Green2025LearningMatch}) where the authors mentioned that $m_{1}$ and $m_{2}$ were the best parametrisation for the higher mass regions. Bias was introduced into this machine-learning pipeline because of the choice of parameters and the distribution they were sampled from, see reference for more information \cite{machinelearningbias}. This showcases that the template bank generated by TemplateGeNN depends on what LearningMatch has learned. 

The dependency of TemplateGeNN on LearningMatch became evident when it was discovered that TemplateGeNN had the ability to highlight the areas in the match manifold that LearningMatch had failed to learn. Throughout the development of LearningMatch, \textit{PyCBC template bank verificator} was used to verify the template banks generated by TemplateGeNN and therefore determine what the Siamese neural network had actually learned. The combination of TemplateGeNN and \textit{PyCBC template bank verificator} meant that the trained LearningMatch models could be scrutinised. This also means that LearningMatch is the bottleneck in this machine-learning pipeline.

LearningMatch has the potential to learn more complex parameter spaces such as non-GR waveforms, precessing and eccentric compact binaries \citep{Sharma2023TemplateBinaries, Wadekar2023AMagnitude, Indik2017StochasticEvents, Lenon2021EccentricExplorer}. This would mean previously computationally expensive template banks that cover large parameter spaces and/or cover a not well-understood parameter metric space can be generated in a reasonable amount of time. Thus, new gravitational-wave search pipelines could be created for gravitational-wave astronomy. TemplateGeNN will also be beneficial for future gravitational-wave detectors, such as LISA \citep{Amaro-Seoane2017} and
Cosmic Explorer \citep{Reitze2020CosmicLIGO}, where templates are computationally expensive
to generate \citep{Lenon2021EccentricExplorer}. However, getting LearningMatch to learn these new match manifolds will be the future challenge. 

\section{Conclusions}
\label{Conclusions}

TemplateGeNN was able to generate a template bank in $\sim 1$ day using a single A100 GPU. The speed-up observed in TemplateGeNN is because the template bank is generated on GPUs rather than CPUs. Current template bank generation pipelines use CPU hardware to compute the match and generate the template. It is worth mentioning that this may not always be the case because there is research making template generation GPU compatible using JAX, see reference for more information \citep{Edwards2023Ripple:Analysis}. 

\textit{PyCBC template bank verificator} was used to determine how well the template bank generated by TemplateGeNN would recover binary black hole mergers injected into Gaussian noise. Out of 7746 injections, 7316 of injections were recovered by the template bank with a fitting factor greater than 0.97, as shown in Figure \ref{fig: Injections Big}. We then focused on the lower mass regions of the template bank and out of 9469 injections, 9416 were recovered with a fitting factor greater than 0.97, as seen in Figure \ref{fig: Injections Small}. In both Figures \ref{fig: Injections Big} and \ref{fig: Injections Small}, the injections that were not recovered by the template bank (i.e. fitting factors less than 0.97) were all in the high mass regions. 

For all the black hole merger injections that were not recovered, LearningMatch was used to determine the maximum match (i.e. fitting factor) between the template bank and the injection, we called this parameter the \textit{LearningMatch Predictive Fitting Factor}. In Figure \ref{fig: LearningMatch Fitting Factor}, \textit{LearningMatch Predictive Fitting Factor} had values greater than 1, which means LearningMatch had predicted match values greater than 1 which is not physically possible. Therefore, we concluded that LearningMatch was responsible for the template bank missing these binary black hole injections. The reason that LearningMatch failed to learn the high mass regions is due to the parametrisation. LearningMatch was trained using the $\lambda_{0}$ parametrisation, compared to the $m_{1}$ and $m_{2}$ parametrisation, because this results in the templates being uniformly distributed in this parameter space. Consequently, in the lower mass regions, the templates get stretched out in $\lambda_{0}$ while in the higher mass regions, the templates get squashed. Therefore, when $\lambda_{0}$ values are taken from a uniform distribution (which is how the training datasets for LearningMatch were created), the higher mass regions are under-covered with data points. This behaviour was also discussed in the original paper (see reference \citep{Green2025LearningMatch}) where the authors mentioned that $m_{1}$ and $m_{2}$ were the best parametrisation for the higher mass regions. Bias was introduced into this machine-learning pipeline because of the choice of parameters and the distribution they were sampled from, see reference for more information \cite{machinelearningbias}. This showcases that the template bank generated by TemplateGeNN depends on what LearningMatch has learned. 

The dependency of TemplateGeNN on LearningMatch became evident when it was discovered that TemplateGeNN had the ability to highlight the areas in the match manifold that LearningMatch had failed to learn. Throughout the development of LearningMatch, \textit{PyCBC template bank verificator} was used to verify the template banks generated by TemplateGeNN and therefore determine what the Siamese neural network had actually learned. The combination of TemplateGeNN and \textit{PyCBC template bank verificator} meant that the trained LearningMatch models could be scrutinised. This also means that LearningMatch is the bottleneck in this machine-learning pipeline.

LearningMatch has the potential to learn more complex parameter spaces such as non-GR waveforms, precessing and eccentric compact binaries \citep{Sharma2023TemplateBinaries, Wadekar2023AMagnitude, Indik2017StochasticEvents, Lenon2021EccentricExplorer}. This would mean previously computationally expensive template banks that cover large parameter spaces and/or cover a not well-understood parameter metric space can be generated in a reasonable amount of time. Thus, new gravitational-wave search pipelines could be created for gravitational-wave astronomy. TemplateGeNN will also be beneficial for future gravitational-wave detectors, such as LISA \citep{Amaro-Seoane2017} and
Cosmic Explorer \citep{Reitze2020CosmicLIGO}, where templates are computationally expensive
to generate \citep{Lenon2021EccentricExplorer}. However, getting LearningMatch to learn these new match manifolds will be the future challenge.


\begin{acknowledgments}
The authors would like to thank Charlie Hoy, Ian Harry and Gareth Cabourn-Davies for their help and useful comments with \textit{PyCBC template bank verificator}. Also, thank you to Laura Nuttall for her useful comments on the paper. Susanna Green was supported by a STFC studentship and the University of Portsmouth. Andrew Lundgren acknowledges the support of UKRI through grants ST/V005715/1 and ST/Y004280/1. The template was generated on the Sciama High Performance Compute (HPC) cluster which is supported by the ICG, SEPNet and the University of Portsmouth. Whilst the template bank was tested using \textit{PyCBC bank verificator} on the computational resources provided by the LIGO Laboratory which are supported by National Science Foundation Grants
PHY-0757058 and PHY-0823459, as well as for additional
computational resources provided by Cardiff University
and funded by STFC grant ST/I006285/1. LIGO is funded by the U.S. National Science Foundation. Virgo is funded by the French
Centre National de Recherche Scientifique (CNRS), the
Italian Istituto Nazionale della Fisica Nucleare (INFN)
and the Dutch Nikhef, with contributions by Polish and
Hungarian institutes. This paper
has been assigned document number LIGO-P2500045.
\end{acknowledgments}

\bibliography{References}

\end{document}